# An Adaptive X-vector Model for Text-independent Speaker Verification

*Bin Gu, Wu Guo, Lirong Dai, Jun Du*

National Engineering Laboratory for Speech and Language Information Processing,
University of Science and Technology of China, Hefei, China
bin2801@mail.ustc.edu.cn, {guowu,lrdai,jundu}@ustc.edu.cn

## Abstract

In this paper, adaptive mechanisms are applied in deep neural network (DNN) training for x-vector-based text-independent speaker verification. First, adaptive convolutional neural networks (ACNNs) are employed in frame-level embedding layers, where the parameters of the convolution filters are adjusted based on the input features. Compared with conventional CNNs, ACNNs have more flexibility in capturing speaker information. Moreover, we replace conventional batch normalization (BN) with adaptive batch normalization (ABN). By dynamically generating the scaling and shifting parameters in BN, ABN adapts models to the acoustic variability arising from various factors such as channel and environmental noises. Finally, we incorporate these two methods to further improve performance. Experiments are carried out on the speaker in the wild (SITW) and VOiCES databases. The results demonstrate that the proposed methods significantly outperform the original x-vector approach.

## 1. Introduction

Speaker verification (SV) is a task to verify a person's claimed identity from speech signals. During the last decade, the i-vector [1] algorithm combined with a probabilistic linear discriminant analysis (PLDA) [2] used for similarity scoring has become a dominant approach for SV.

This paradigm has been improved by incorporating a deep neural network (DNN) to extract speaker representations, which are named the x-vector [4] or d-vector [30] in the SV field. In most of these DNN-based systems, several frame-level layers are stacked to deal with a local short span of acoustic features to obtain more effective high-level representations. These layers can be modeled by a time-delay neural network (TDNN) [3, 4], convolutional neural network (CNN) [5] or long short-term memory network (LSTM) [6, 7]. Then, a pooling layer maps all frames of the input utterance into a fixed-dimensionality vector, and speaker embedding is generated from the following stacked fully connected layers. Average pooling, max pooling [8] and statistical pooling [3] are widely used in pooling layers. Some researchers have also employed the attention mechanism [9] and gating mechanism [10, 11] in the pooling layer. By providing different frame weights, these methods can capture more expressive speaker characteristics. Such DNN embedding systems have become the current state-of-the-art systems in most public benchmarks.

Speech signals are easily corrupted by various factors, such as emotions, channels, and environmental noises. How to extract robust speaker embeddings is one of the principal interests of SV. The data augmentation technique [4, 12, 13] is the most straightforward way to solve this problem. The systems can achieve better performance by constructing additional training samples using expert knowledge or extra data sources. Another choice is applying the adversarial training strategy in the speaker characteristics modeling process. Through weakening the ability to discriminate the environment types, SNRs [14] or other relative information [15, 16] in a speech, the speaker embedding extractor generates more robust speaker representations.

Recently, an adaptive convolution neural network (ACNN) has proven to be useful for the natural language processing (NLP) tasks [17, 18] and the computer vision (CV) tasks [19]. Unlike traditional convolutions that use the same set of filters regardless of different inputs, adaptive convolution employs adaptively generated convolutional filters that are conditioned on inputs. Similar to these works, dynamic layer normalization (DLN) [20] and adaptive batch normalization (ABN) [21] are proposed for adaptive neural acoustic modeling in speech recognition. The parameters in the normalization layer are substituted with learned functions, the outputs from which are then used as normalization parameters. In these studies, an adaptive mechanism gives stronger flexibility to networks and allows networks to utilize the information inputs contained.

In this paper, we investigate the abovementioned adaptive learning methods for robust speaker embedding extraction. More specifically, the ACNN and ABN are employed in the frame-level layers for extracting more expressive feature representations. In addition, we incorporate these two methods into an x-vector network to further improve performance. To the best of our knowledge, this study is the first to employ input-aware methods to extract robust speaker embeddings. We evaluate our experiments on the SITW and VOiCES datasets. The experimental results show that the two methods can both achieve better performance than the original x-vector approach, and the appropriate integration of the methods can further improve performance.

The remainder of this paper is organized as follows. Section 2 gives an introduction to our x-vector baseline. Section 3 describes the proposed input-aware model in detail. Then, the experimental setup, the results and the analysis are presented in section 4. Finally, the conclusions are given in section 5.

## 2. Baseline network architecture

The network architecture of our x-vector baseline system is the same as that described in [4]. As depicted in Figure 1, the x-vector baseline consists of three time-delay frame-level layers, two more frame-level layers without time delay, a pooling layer that converts the variable-length frame-level representations into a single fixed-length vector and, finally, two utterance-level layers followed by the output layer.

As we know, the TDNNs in the frame-level layers could be implemented as 1-D convolutional neural networks (1-D CNNs), where the filters slide along the time axis. The mean and standard deviation of the final frame-level output vectors are calculated and then concatenated together for the pooling layer. All activations in the network are rectified linear units (ReLUs). Batch normalization is used on the activations from all layers except the output layer. The output layer computes an affine transform of its input and then transforms the outputs using softmax. The network is trained to predict the correct speaker labels with cross entropy (CE) loss. Once the DNN is trained, the speaker embeddings are extracted from the layer right after pooling.

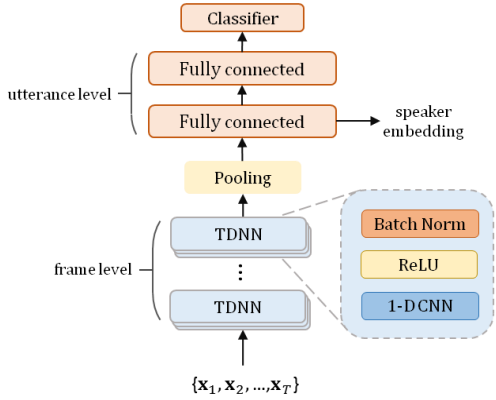

Figure 1 Baseline network architecture

# 3. Adaptive X-vector Model

This section introduces the proposed ACNN and ABN. We explain how the parameters are generated and applied to existing framework.

## 3.1. Adaptive convolution neural network

Figure 2 schematically shows the overall architecture of our ACNN. Attentive statistic pooling is used to encapsulate the variable size input into a fixed size context vector at first, and this vector adjusts the convolution parameters by determining the weights of the component filters and biases. The final convolutional parameters are linear regressions of these components.

Suppose $\mathbf{h}_t^l$ is the hidden representation in the $l^{th}$ layer. The attention mechanism is applied first. The value vectors $\mathbf{e}_t$ and the attention weights $\alpha_t$ are calculated as follows:

$$\mathbf{e}_t = \mathbf{h}_t^l * \mathbf{W}_e + \mathbf{b}_e$$
$$\alpha_t = \mathbf{v}^T \tanh(\mathbf{h}_t^l * \mathbf{W}_\alpha + \mathbf{b}_\alpha) \qquad (1)$$

where $\mathbf{W}_e$ and $\mathbf{W}_\alpha$ are the convolution parameters, while $\mathbf{b}_e$ and $\mathbf{b}_\alpha$ are the bias parameters. $\mathbf{v}$ is a vector that converts the hidden vector to a scalar value. Then, we generate a context vector by leveraging the statistical information inherent in the weighted value vectors $\mathbf{e}_t$.

$$\boldsymbol{\mu} = \sum_t \alpha_t \mathbf{e}_t$$
$$\boldsymbol{\sigma} = \sqrt{\sum_t \alpha_t \mathbf{e}_t \odot \mathbf{e}_t - \boldsymbol{\mu} \odot \boldsymbol{\mu}}$$
$$\mathbf{c}_{acnn} = [\boldsymbol{\mu}, \boldsymbol{\sigma}] \qquad (2)$$

Finally, we concatenate $\mathbf{u}$ and $\boldsymbol{\sigma}$ as the context vector $\mathbf{c}_{acnn}$ to generate the convolutional parameters by a linear combination of components from a parameter pool.

$$\boldsymbol{\beta} = [\beta_1, ..., \beta_N] = \mathbf{W}_\beta \mathbf{c}_{acnn} + \mathbf{b}_\beta$$
$$\tilde{\mathbf{W}} = \sum_{i=1}^N \beta_i \mathbf{W}_i$$
$$\tilde{\mathbf{b}} = \sum_{i=1}^N \beta_i \mathbf{b}_i \qquad (3)$$

where $\mathbf{W}_\beta$ and $\mathbf{b}_\beta$ are trainable parameters to generate the weight vector $\boldsymbol{\beta}$ and the parameters $\{\mathbf{W}_i\}_{i=1...N}$ and $\{\mathbf{b}_i\}_{i=1...N}$ of the component filters can also be trained through a typical backpropagation algorithm.

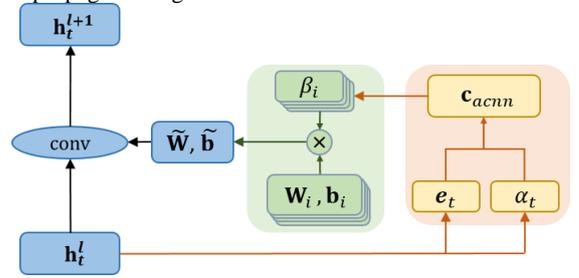

Figure 2 Structure of the proposed ACNN

Once we obtain the convolutional filter $\tilde{\mathbf{W}}$ and bias $\tilde{\mathbf{b}}$, a conventional convolution operation is applied to the inputs as follows:

$$\mathbf{h}_t^{l+1} = f(\mathbf{h}_t^l * \tilde{\mathbf{W}} + \tilde{\mathbf{b}}) \qquad (4)$$

where $f$ is a nonlinear function and is usually composed of an activation function and batch normalization.

## 3.2. Adaptive batch normalization

The main difference between the ABN and BN is that the scaling and shifting parameters are dynamically generated for different inputs in the ABN while they are fixed for all the inputs in the BN in the testing procedure. The main procedure of the ABN is similar to that of the ACNN. First, the weighted context vector $\mathbf{c}_{abn}$ can be calculated as follows:

$$\mathbf{e}_t = \tanh(\mathbf{W}_e * \mathbf{h}_t^l + \mathbf{b}_e)$$
$$\alpha_t = \frac{\exp(mean(\mathbf{e}_t))}{\sum_i \exp(mean(\mathbf{e}_i))}$$
$$\mathbf{c}_{abn} = \sum_t \alpha_t \mathbf{e}_t \qquad (5)$$

where $\mathbf{W}_e$ and $\mathbf{b}_e$ are trainable parameters. The mean of all the elements in $\mathbf{e}_t$, which is the nonlinear transformed low-dimension vector of input $\mathbf{h}_t^l$, is used to measure the importance of each frame, and the weighted sum of $\mathbf{e}_t$ is used as the context vector $\mathbf{c}_{abn}$. Then, the scaling $\mathbf{\gamma}$ and shifting $\mathbf{\beta}$ parameters are generated from $\mathbf{c}_{abn}$.

$$\mathbf{\gamma} = \mathbf{W}_\gamma \mathbf{c}_{abn} + \mathbf{b}_\gamma$$
$$\mathbf{\beta} = \mathbf{W}_\beta \mathbf{c}_{abn} + \mathbf{b}_\beta \tag{6}$$

where $\mathbf{W}_\gamma$, $\mathbf{W}_\beta$, $\mathbf{b}_\beta$ and $\mathbf{b}_\gamma$ are all trainable parameters. Finally, standard batch normalization [22] is employed with the generated parameters.

## 4. Experiments and Discussion

### 4.1. Data set and evaluation metrics

All experiments are conducted on the SITW and VOiCES datasets. For the SITW dataset [23], there are two standard datasets for testing: dev. core and eval. core. We use both sets to conduct the experiments. The VoxCeleb database [24], including the VoxCeleb1 and VoxCeleb2, is used for training. Since a few speakers are included in both the SITW and VoxCeleb datasets, these speakers are removed from the training dataset.

The VOiCES dataset for the speaker verification task is described in the "VOiCES from a Distance Challenge 2019" [25]. The development dataset contains 15,904 noisy and far-field speech segments from 196 speakers. The evaluation set consists of 11,392 distant recordings from different microphone types and different rooms, both of which could be more challenging than those featured in the development set.

Due to the background noise, reverberation, laughter and acoustic artifacts contained in speech data, the data augmentation techniques described in [4], including adding additive noise and reverberation data, are applied to improve the robustness of the system. Because there is a possibility that there will be an overlap between the MUSAN [26], which is a publicly available augmentation dataset, and the VOiCES dataset, babble noise is not created for augmentation. In summary, there are a total of 2,236,567 recordings from 7185 speakers for training, including approximately 1,000,000 randomly selected augmented recordings. Note that the training data for VOiCES are consistent with those for the SITW dataset.

The results are reported in terms of three metrics: the equal error rate (EER), the minimum of the normalized detection cost function (minDCF) and the actual detection cost function (actDCF). The minDCF has two settings: one with a prior target probability $P_{tar}$ set to 0.01 (DCF($10^{-2}$)) and the other with a $P_{tar}$ set to 0.001 (DCF($10^{-3}$)).

### 4.2. Features

We select 30-dimensional MFCC features containing delta and delta-delta coefficients as the input acoustic features. Each MFCC feature is extracted from the speech signal with a 25ms window and a 10ms frame shift. Each feature is mean-normalized over a 3 s sliding window, and energy-based VAD is employed to filter out non-speech frames. The acoustic features are randomly cropped to lengths of 2-4 s, and 128 utterances with the same duration are grouped into a mini-batch. Data processing is implemented with the Kaldi toolkit [27].

### 4.3. Model configuration

All neural networks are implemented using the TensorFlow toolkit [28]. The network is optimized using the Adam optimizer, and the learning rate gradually decreases from 1e-3 to 1e-4. If not specified, all of the setups are the same as the baseline system. Other configurations of each system are listed as follows:

**x-vector**: This is a deep embedding learning baseline system. Only the fifth hidden layer has 1536 nodes, while the other layers have 512 nodes. The kernel sizes of the first five layers are 5, 3, 3, 1 and 1, while the dilation rates are set to 1, 2, 3, 1 and 1 respectively. The same type of L2 weight decay and batch normalization as described in [29] are used in the baseline system to prevent overfitting.

**ACNN**: In this system, the ACNN is only applied in the fourth frame-level layer. Such a setup can achieve satisfactory results with a minimum increase in parameters. The hidden dimensions of both $\mathbf{W}_e$ and $\mathbf{W}_\alpha$ in Eq. (1) are set to 256. The number of component filters N in Eq. (3) is chosen to be 4.

**ABN**: All frame-level layers employ the ABN in this system. The hidden dimension of $\mathbf{W}_e$ in Eq. (5) is set to 256. Note that the utterance-level layers use the conventional BN, and other setups are exactly the same as the baseline system.

**ACNN&ABN**: Both the ACNN and ABN are employed in this system. The ACNN is employed in the fourth layer, while the ABN is used in the remaining frame-level layers. The setup is consistent with the abovementioned **ACNN** and **ABN** systems.

**Fusion**: The complementarity between the above two different adaptive learning methods at the score level is also investigated here. We only report the results using the score fusion of the ACNN and ABN with equal weights.

### 4.4. PLDA Backend

The DNN embeddings are centered using the training set and are projected to a low-dimensional space using LDA at first. The dimensions of the x-vectors are reduced to 100 for both datasets. After length normalization, we select the longest 200,000 recordings from the training set to train the PLDA backend. The backend classifier is implemented with the Kaldi toolkit.

### 4.5. Results and analysis

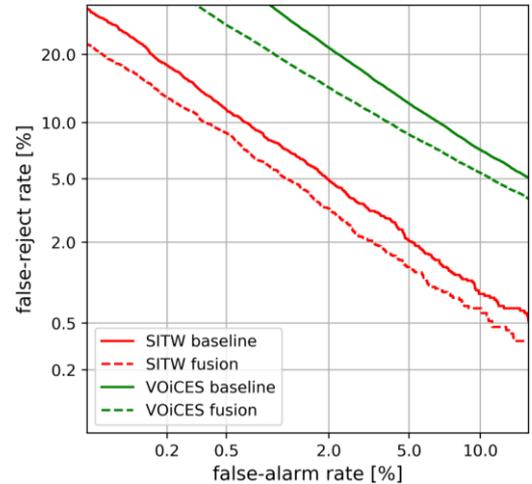

Figure 4 *DET curve comparison for the evaluation set of SITW and VOiCES*

Table 1 presents the results of different systems on the SITW and VOiCES datasets. It can be observed that the system applying either the ACNN or ABN outperforms the x-vector baseline system. On the SITW dataset, these two systems can both improve the baseline by approximately 10% ~ 19% for all evaluation metrics. For the VOiCES dataset, the ACNN achieves at most 21% relative improvements over the baseline in terms of minDCF and actDCF, while the ABN achieves at most 17% relative improvements in terms of EER. The ACNN&ABN system can obtain some further performance improvement over the ACNN and ABN systems, especially on the SITW dataset.

Table 1 *Results of different systems on the SITW and VOiCES datasets. DCF2, DCF3 and aDCF refer to DCF(10⁻²), DCF(10⁻³) and actDCF, respectively. **Boldface** values are the best results. Impr denotes the relative improvement of the best results with respect to the baseline system.*

| System | SITW | | | | | | VOiCES | | | | | |
|---|---|---|---|---|---|---|---|---|---|---|---|---|
| | Dev | | | Eval | | | Dev | | | Eval | | |
| | EER | DCF2 | DCF3 | EER | DCF2 | DCF3 | EER | DCF2 | aDCF | EER | DCF2 | aDCF |
| x-vector | 2.88 | 0.2956 | 0.4752 | 3.280 | 0.3063 | 0.4974 | 3.44 | 0.3952 | 0.4925 | 8.339 | 0.6203 | 0.7299 |
| ACNN | 2.54 | 0.2389 | 0.4126 | 2.734 | 0.2824 | 0.4430 | 3.25 | 0.3346 | 0.3663 | 7.568 | 0.5553 | 0.5800 |
| ABN | 2.35 | 0.2444 | 0.4110 | 2.898 | 0.2765 | 0.4380 | 2.87 | 0.3206 | 0.4060 | 7.469 | 0.5709 | 0.6167 |
| ACNN&ABN | 2.35 | 0.2317 | **0.3693** | **2.542** | 0.2687 | 0.4233 | 2.73 | 0.3263 | 0.4291 | 7.121 | 0.5676 | 0.6090 |
| Fusion | **2.12** | **0.2264** | 0.3768 | 2.597 | **0.2650** | **0.4106** | 2.71 | **0.2841** | **0.3060** | **7.029** | **0.5114** | **0.5182** |
| Impr.(%) | 26 | 23 | 21 | 21 | 13 | 17 | 21 | 28 | 38 | 16 | 18 | 29 |

Table 2 *Comparison results of the proposed ACNN system using different setups. Except for the parameter N, the rest of the setup is consistent with the ACNN system described in section 4.3*

| System | SITW | | | | | | VOiCES | | | | | |
|---|---|---|---|---|---|---|---|---|---|---|---|---|
| | Dev | | | Eval | | | Dev | | | Eval | | |
| | EER | DCF2 | DCF3 | EER | DCF2 | DCF3 | EER | mDCF | aDCF | EER | mDCF | aDCF |
| ACNN(N=2) | **2.347** | **0.2379** | **0.3995** | 2.816 | 0.2871 | 0.4457 | 3.253 | 0.3557 | 0.4083 | 7.867 | 0.5677 | 0.6144 |
| ACNN(N=4) | 2.541 | 0.2389 | 0.4126 | **2.734** | **0.2824** | **0.4430** | 3.254 | **0.3346** | **0.3663** | **7.568** | **0.5553** | **0.5800** |
| ACNN(N=6) | 2.542 | 0.2595 | 0.4060 | 3.007 | 0.2828 | 0.4562 | **2.991** | 0.3752 | 0.4764 | 7.952 | 0.6158 | 0.7067 |
| ACNN(N=8) | 2.580 | 0.2547 | 0.4234 | 2.816 | 0.2855 | 0.4558 | 3.239 | 0.3698 | 0.4321 | 7.741 | 0.6126 | 0.6847 |

Among all of the above systems, the fused system achieves the best performance especially in terms of the actDCF, which improves over the baseline by nearly 38% and 29% on the development set and the evaluation set of VOiCES, respectively. Figure.3 depicts the detection error trade-off (DET) curves of the baseline and the fusion systems, and obvious improvements can be observed.

Note that we only employ the ACNN in the fourth frame-level layer for two reasons. In the ACNN, the number of parameters is usually several times higher than the number of parameters in the CNN because each component filter is the same size as that in the CNN. The fourth frame-level layer has the minimum convolution kernel size and hidden dimension in our systems. Applying the ACNN in such a layer only causes an approximate 9% increase in parameters, while applying it in any other layer causes at least a 27% increase in parameters with respect to the baseline. On the other hand, the three bottom frame-level layers model long-term temporal dependencies with time delay and the frame-level feature representations are not high-level enough to reflect all kinds of information in high-dimensional abstract space.

The hyperparameter $N$ in Eq. (3) controls the number of component filters. The experimental results with different $N$ are listed in Table 2. There is a larger gap between the development set and evaluation set with $N=2$. This means that too few component filters cannot guarantee the generalizability of the model. Generally the system with $N=4$ achieves the best performance. Therefore, $N=4$ is set in our experiments.

To test the effectiveness of the proposed ACNN and ABN in different speech environments, we select clean utterances and degraded utterances from the SITW dataset that are naturally degraded with noise, compression and reverberations. The trials were divided into 4 groups according to the speech quality and marked as "clean", "noise", "codec" and "Reverb" respectively. The results in Table 3 and 4 show that both the ABN and ACNN achieve significant improvements under all conditions. This comparison demonstrates that the proposed algorithms are robust to environment types and speech quality.

Table 3 *Comparison results of the SITW development set under different conditions*

| System | Clean | Noise | Codec | Reverb |
|---|---|---|---|---|
| x-vector | 3.226 | 2.675 | 3.085 | 2.437 |
| ACNN | 3.225 | 2.171 | 2.587 | 2.366 |
| ABN | **1.613** | **1.918** | 2.586 | 2.079 |
| ACNN&ABN | 2.419 | 2.221 | **2.447** | **1.935** |

Table 4 *Comparison results of the SITW evaluation set under different conditions*

| System | Clean | Noise | Codec | Reverb |
|---|---|---|---|---|
| x-vector | 4.819 | 2.833 | 3.816 | 2.798 |
| ACNN | 4.418 | 2.366 | **2.961** | 2.467 |
| ABN | **4.016** | 2.617 | 3.092 | 2.468 |
| ABN&ACNN | **4.016** | **2.225** | 3.355 | **2.323** |

## 5. Conclusions

In this study, we employ adaptive mechanisms in the DNN embedding system to adaptively utilize the information of inputs. More specifically, the ACNN is introduced into the frame-level layers where the output representations are carefully modulated by adaptively estimating the convolution filters and biases. Such a mechanism helps to obtain more expressive features. Furthermore, the batch normalization layer is enhanced by dynamically generating the shifting and scaling parameters. The experimental results demonstrate that the adaptive mechanics outperform the conventional x-vector baseline. The proposed two methods have obvious complementarity with each other especially at the score level.

In our future studies, we will continue to focus on the use of adaptive mechanisms and investigate other useful deep learning strategies to enhance our proposed methods for x-vector based speaker verification.

## 6. Acknowledgements

This work was partially funded by the National Natural Science Foundation of China (Grant No. U1836219) and the National Key Research and Development Program of China (Grant No. 2016YFB100 1303).